\RequirePackage[normalem]{ulem} 
\RequirePackage{color}\definecolor{RED}{rgb}{1,0,0}\definecolor{BLUE}{rgb}{0,0,1} 

\documentclass[12pt,usenatbib]{mn2e}
\usepackage{url,ulem,times,graphicx,amsmath,amsfonts,amssymb,color,epsfig,epstopdf}
\usepackage{graphics}
\usepackage{epsf}
\usepackage{bm}
\usepackage{color}
\usepackage{rotating}
\usepackage[T1]{fontenc}
\usepackage{ae,aecompl}
 \usepackage{array,multirow}
\usepackage[percent]{overpic}

\usepackage{hyperref}
\hypersetup{
   colorlinks=true,
   urlcolor=blue,
   citecolor=blue,
   pdfborder= 0 0 0
}

\def\ethree{${\bf e}_{3}$}

\begin{document}

\title[Bisous and V-web filaments]{Filaments from the galaxy distribution and from the velocity field in the local universe}
\author[Libeskind et al]
{Noam I. Libeskind$^{1}$, Elmo Tempel$^{2}$, Yehuda Hoffman$^{3}$,  R. Brent Tully$^{4}$ \newauthor 
H\'{e}l\`{e}ne Courtois$^{5}$\\
$^1$Leibniz-Institut f\"ur Astrophysik Potsdam (AIP), An der Sternwarte 16, D-14482 Potsdam, Germany\\
$^{2}$Tartu Observatory, Observatooriumi~1, 61602 T\~oravere, Estonia\\ 
 $^3$Racah Institute of Physics, Hebrew University, Jerusalem 91904, Israel\\
 $^4$Institute for Astronomy (IFA), University of Hawaii, 2680 Woodlawn Drive, HI 96822, US\\
 $^5$University of Lyon; UCB Lyon 1/CNRS/IN2P3; IPN Lyon, France\\
  }

\date{Accepted --- . Received ---; in original form ---}
\pagerange{\pageref{firstpage}--\pageref{lastpage}} \pubyear{2015}
\maketitle
\label{firstpage}

 \begin{abstract}
The cosmic web that characterizes the large-scale structure of the Universe can be quantified by a variety of methods. For example, large redshift surveys can be used in combination with point process algorithms to extract long curvi-linear filaments in the galaxy distribution. Alternatively, given a full 3D reconstruction of the velocity field, kinematic techniques can be used to decompose the web into voids, sheets, filaments and knots.  In this paper we look at how two such algorithms -- the Bisous model and the velocity shear web -- compare   with each other in the local Universe (within 100~Mpc), finding good agreement. This is both remarkable and comforting, given that the two methods are radically different in ideology and applied to completely independent and different data sets. Unsurprisingly, the methods are in better agreement when applied to unbiased and complete data sets, like cosmological simulations, than when applied to observational samples. We conclude that more observational data is needed to improve on these methods, but that both methods are most likely properly tracing the underlying distribution of matter in the Universe.
\end{abstract}

\section{Introduction}
\label{section:intro} 

That galaxies are not homogeneously and uniformly distributed in the Universe has been known at least since  \citet[][]{1923MNRAS..83..147R} stated that ``... there is very definite evidence for a band [of galaxies], fairly widespread ...''  \citep[see also][]{1847QB3.H52........}, a feature later confirmed and named the ``supergalactic plane''. In modern terms, the anisotropic collapse of density perturbations into large plane-like pancakes was predicted in the seminal work of \cite{1970A&A.....5...84Z}. The concepts of filaments in the galaxy distribution was first suggested by \cite{1978MNRAS.185..357J}, while \cite{1996Natur.380..603B} argued that the large-scale distribution of matter constitutes a network known as the cosmic web.

Since then, many researchers have sought to quantify the cosmic web using a variety of techniques to gain insight into the relationship between galaxies and their environment. Often simulations are used as a testbed, since the full 6~dimensional phase space of the matter distribution is available. Sophisticated algorithms that attempt to capture the cosmic web's multi-scale nature \cite[e.g.][]{2013MNRAS.429.1286C}, or that are based on shell crossing and phase-space folding \citep{2012ApJ...754..126F,2012PhRvD..85h3005S,2012MNRAS.427...61A} have proven useful to understand how the cosmic web evolves \citep[see also][]{2007A&A...474..315A,2010ApJ...723..364A}. Other methods, for example based on the Hessian of the tidal \citep{2007MNRAS.375..489H,2009MNRAS.396.1815F}  or shear tensor \citep{2012MNRAS.425.2049H,2014MNRAS.441.1974L} have proven useful to study aspects of haloes such as their mass function \citep{2015MNRAS.446.1458M} or spin \citep{2012MNRAS.421L.137L,2012MNRAS.427.3320C}.

The construction of the cosmic web from astronomical data represent a considerable challenge. A partial list of obstacles to overcome includes magnitude limits, volume limits, biases of multiple sorts, incompleteness, redshift-space distortions, obscuration by the Galactic disc and sparse sampling. Nevertheless, various techniques  to find filaments in observational samples have been devised \citep[e.g.][among others]{2008ApJ...672L...1S,2011MNRAS.414..384S,2010MNRAS.409..156B,2010A&A...510A..38S,2014MNRAS.438..177A,2015MNRAS.450.1999H}. The correspondence between numerical and observational methods remains unclear and further study is clearly necessary.

As an attempt to resolve this ambiguity, \cite{2014MNRAS.437L..11T} examined the alignment between cosmic filaments detected by a point process with those detected by exploiting the velocity shear tensor, using an $N$-body simulation. They found that indeed these methods agree to a large extent. The work of  \citet{2014MNRAS.437L..11T}  is extended here from simulations and applied to two data sets representing the large-scale structure of the local (<100~Mpc) Universe.  We compare the filaments constructed from the Bisous method applied to the 2MRS \citep{2012ApJS..199...26H, 2011MNRAS.416.2840L} catalogue of galaxies, to those obtained by the velocity shear method  \citep{2012MNRAS.425.2049H} from the Wiener Filter reconstruction of the Cosmicflows-2 \citep[CF2,][]{2013AJ....146...86T} data base.

The agreement that \citet{2014MNRAS.437L..11T} found in simulations indicated that these two methods trace the same cosmic structures. However a comparison between these two methods applied to simulations is bound to give a better agreement than when applied to observational data. For one, \cite{2014MNRAS.437L..11T} had the full density and velocity field at their disposal from which to compute the shear field and Bisous filaments; when using observations only magnitude limited, partially obstructed, galaxy redshift surveys are available for the latter and sparse, incomplete and inhomogeneous catalogues of radial velocities are available for the former. In \cite{2014MNRAS.437L..11T}, all dark matter haloes (above a minimum mass cut) in the computational box were used as proxies of the galaxy distribution returning fairly complete filaments throughout the simulated volume; in observations, galaxies obtained according to the survey's selection function is used for this purpose. Since simulations offer the full 6D phase space information of the density and velocity fields to high accuracy, the shear tensor can be constructed on very small scales. When dealing with the sparse sampling of the 4D (position and radial velocity) phase space information of observations, the resolution of the shear tensor computation is significantly worse. It is unclear whether these two methods can agree at all on observational data, given the long list of limitations.

\section{Method}
\subsection{V-web filaments from Cosmicflows-2}
\label{sec:WF}
The CF2 data \citep{2013AJ....146...86T} are a comprehensive compilation of the universe's {\it radial} peculiar velocity field, within around 100~Mpc of the Milky Way. The radial peculiar velocity field can be used to reconstruct the full 3D velocity and density fields by means of a Wiener Filtering algorithm \citep{2012ApJ...744...43C}. We refer the interested reader to the important papers describing Wiener Filtering and constrained realizations for details \citep{1991ApJ...380L...5H,1994ApJ...423L..93L,1995ApJ...449..446Z,1999ApJ...520..413Z,1995MNRAS.272..885F} and explain the method only schematically here.

The Wiener Filter is essentially a linear minimum variance solution to recover an underlying, continuous field given a set of noisy, sparse, and incomplete data, such as the CF2 survey. The technique assumes a prior for the form of the covariance matrix of the desired field. In this case, the prior is derived from the assumption of a $\Lambda$CDM universe where structures form out of Gaussian random fields. The Wiener Filtering algorithm takes as an input the measured radial velocity field and returns a reconstructed 3D (linear) velocity field which can be used to quantify the cosmic web kinematically. 
There is an implicit smoothing in the Wiener Filter which is adaptive and inversely proportional to the quality of the data, namely the sampling and magnitude of the observational errors. Where the observational errors are large or where the sampling is poor, the Wiener Filter has a large effective smoothing. In practice, the reconstructed velocity and density fields are attenuated with the depth, since observational errors are large and sampling is poor at great distances. In the limit of vanishing data the Wiener Filter converges to the null fields. Nearby the effective resolution is $\sim 5$~Mpc.

A kinematic classification can be made by examining the (reconstructed) velocity field, as described in \cite{2012MNRAS.425.2049H}. A regular ($256^{3}$) grid is placed on the velocity field and the shear tensor (similar in form to the ``stress-strain'' tensor used in mechanics) is computed and diagonalized to return the principal components, known as the eigenvectors and corresponding eigenvalues. These describe the directions along which matter is principally collapsing (or expanding). The number of eigenvalues above an (arbitrary) threshold can be used to classify a region as belonging to a void, sheet, filament, or knot. Typically the value for the threshold is chosen to recover the visual impression of the cosmic web \citep[e.g. see][]{2009MNRAS.396.1815F}, but can also often taken to be null such that positive eigenvalues correspond to axes along which compression occurs while negative eigenvalues imply expansion. In this work, we chose the latter threshold value of zero.

With such a threshold, regions with 3 positive eigenvalues ($+,+,+$) correspond to knots, 2 positive eigenvalues ($+,+,-$) to filament, 1 positive eigenvalue ($+,-,-$) to sheets and 0 positive eigenvalues ($-,-,-$) to voids. Eigenvalues and corresponding eigenvectors are conventionally denoted as ($\lambda_{i},~{\bf e}_{i}$) with $i=1,2,3$, respectively. In the case of filaments, one has a single negative eigenvalue ($\lambda_{3}$); the corresponding eigenvector (\ethree), defines the ``spine'' of the filament. Its direction can be measured against the direction of filaments computed with the Bisous method, described below.

\subsection{Bisous filaments from marked point process}
\label{sec:bisous}
Filaments in the local Universe can be found by linking galaxies from the 2MRS survey \citep{2012ApJS..199...26H, 2011MNRAS.416.2840L} together in a curvi-linear fashion. The so-called ``Bisous model''  applied a marked point process with interactions \citep{Stoica:05} to the 3D spatial distribution of galaxies. The method provides a quantitative classification that agrees well with a visual impression of the filamentary nature of the cosmic web. More details regarding the Bisous model can be found in \citet{2007JRSSC..56....1S,2010A&A...510A..38S} and \citet{2014MNRAS.438.3465T}. A brief, albeit simple summary is provided below.

The Bisous model approximates the filamentary web by randomly placing small cylindrical segments on the galaxy distribution, and finding those cylinders which best ``fit'' the local, linear distribution of galaxies. The model assumes that locally the galaxy distribution can be approximately described by a series of relatively small cylinders, which can be combined to trace a filament if the neighbouring cylinders are oriented similarly. One of the advantages of such ``point processes'' is that they rely directly on the observed galaxy positions, and does not require any additional post-processing of the data for example the computation of continuous density field via smoothing kernels.

The solution provided by the Bisous model is stochastic and thus relies on the generation of  Markov-chain Monte Carlo (MCMC). It is thus expected that there is some variation in the detected filamentary patterns for different MCMC runs of the model. However, thanks to the stochastic nature of the Bisous model, we gain a morphological and a statistical characterization of the filamentary pattern simultaneously.

The only free parameter in the model is the approximate radius of the filamentary cylinder. Once this is specific, the algorithm returns the filament detection probability field as well as the filament orientation field. Based on these two fields, a filament catalogue can be built in which every filament is represented a specified axis, known as its spine.

The spine detection follows two ideas. First, filament spines are required to be located where the algorithm specifies the filament detection probability is highest. In practice this is done by algorithmic examination of  probability maps. Second, in these regions of high probability for the filamentary network, the spines are oriented along the orientation field of the filamentary network. See \citet{2014MNRAS.438.3465T} for more details of the procedure.

The filament scale in this study is chosen to be \mbox{$\sim0.7$~Mpc}. This scale and all other parameters in Bisous model are exactly the same as in \citet{2014MNRAS.438.3465T}. The chosen filament scale corresponds to the scale of galaxy groups/clusters and the detected filaments are the bridges between them.

\section{Results}
\begin{figure}
 \includegraphics[width=20pc]{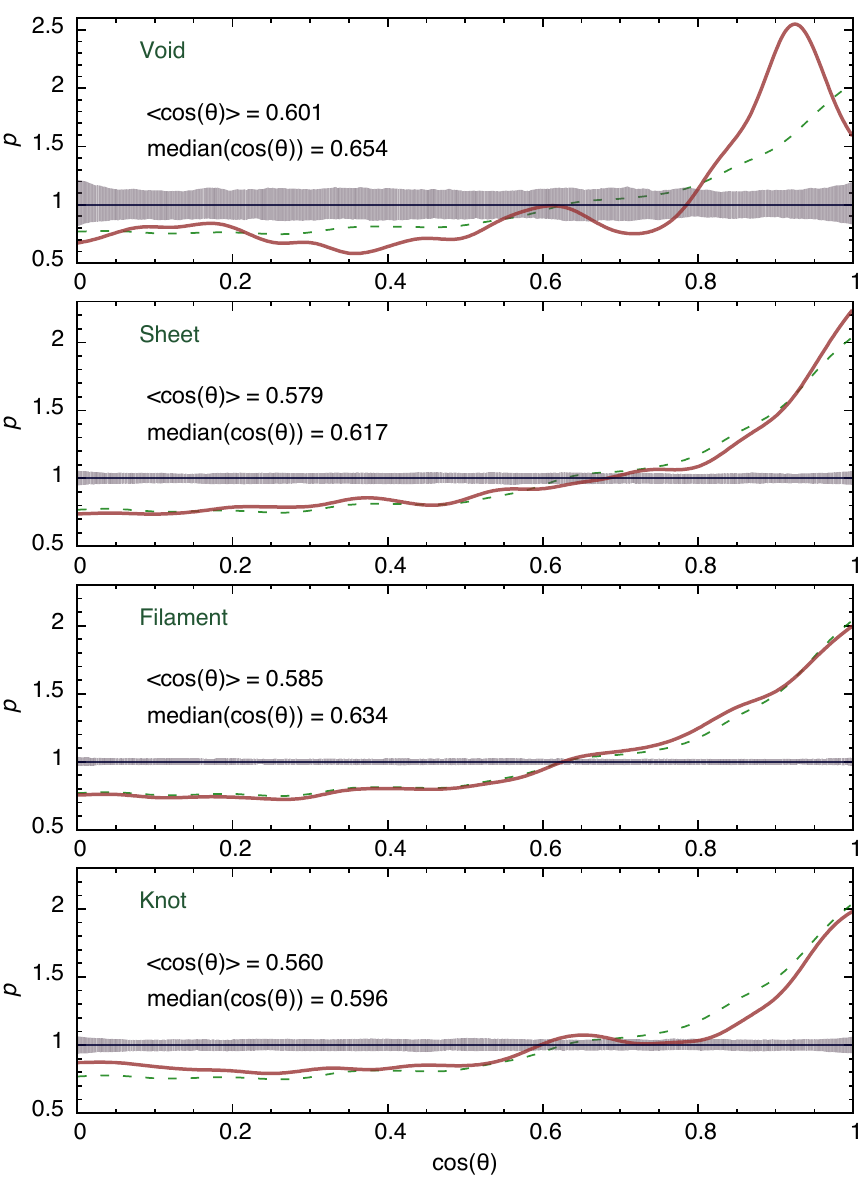}
 \caption{The probability distribution of the cosine of the angle formed between Bisous filaments and \ethree, the axis of slowest collapse or fastest expansion computed from the velocity shear tensor. In each panel, the green dashed line indicates the alignment for all Bisous filaments. The Bisous filaments are then subdivided by V-web classification into voids, sheets, filaments, and knots (top to bottom) and the probability distribution in each web-type is shown by the red solid line in each panel. The mean and median $\cos(\theta)$ are also shown. If the two directions were randomly oriented with respect to each other the probability distribution would be uniform at unity. 95\% confidence region of a random distribution is shown by the grey shaded region. All alignments are statistically very strong.}
 \label{fig:pd}
\end{figure}
\begin{figure}
 \includegraphics[width=20pc]{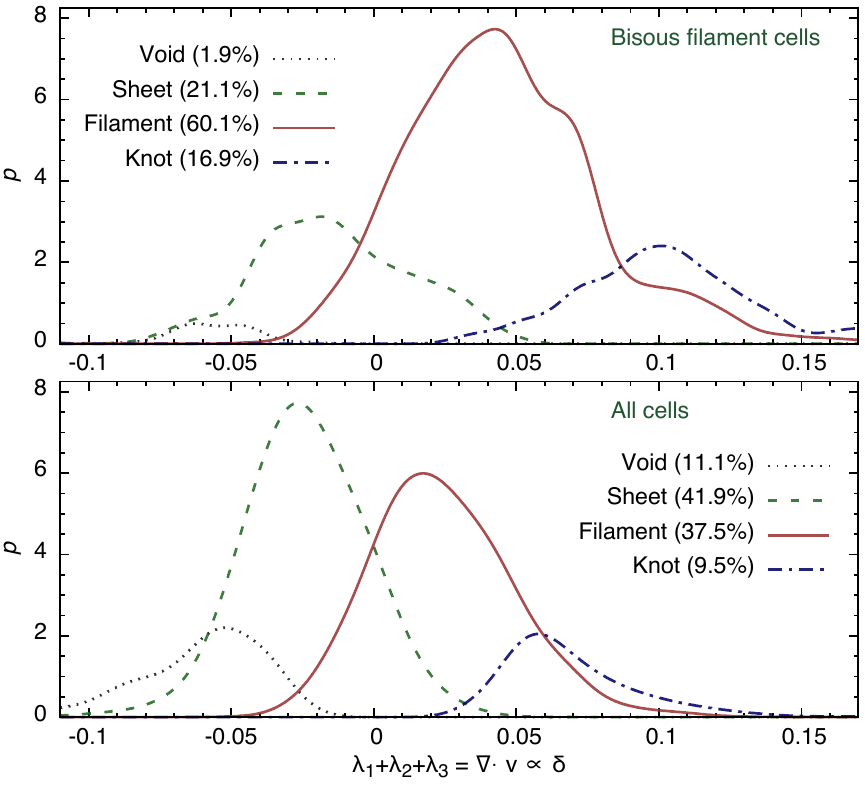}
 \caption{The sum of the shear tensor's eigenvalues is the divergence of the velocity field ($\lambda_{\rm 1}+\lambda_{\rm 2}+\lambda_{\rm 3} =\nabla\cdot v$ ) which on the linear scales of the Wiener Filter is proportional to the over-density ($\nabla\cdot v \propto \delta$). The distribution of the over-density in the four web types (voids, black dotted; sheets, green dashed; filaments, red solid; knots, blue dot-dashed) are shown here. {\it Top panel} shows this only for regions (cells) that are coincident with Bisous filaments, while the {\it bottom panel} shows this quantity for the full reconstructed CF2 volume.}
 \label{fig:d}
\end{figure}

After computing the Bisous filament catalogue from 2MRS and the direction (\ethree) corresponding to the spine of filaments extracted from the Wiener Filter reconstruction of the velocity field, we are able to measure the alignment of the two. Since the Wiener Filter and reconstructed velocity field performs best where the sampling is greatest (and attenuates to the mean field when the signal is lost), we restrict our analysis to regions of the universe that are within 100~Mpc. The fraction of the volume (within 100~Mpc) ascribed to knots, filaments, sheets and voids within this distance is 9.5, 37.5, 41.9, 11.1\%, respectively (shown also in Fig.~\ref{fig:d}, bottom panel).

 In practice, we find the locations of filaments in the Bisous catalogue of 2MRS galaxies first. Then, at the location of each filament, we select the corresponding \ethree~eigenvector from the reconstructed velocity field. The two are dotted and the distribution of the (cosine) of the angle between them is shown in Fig.~\ref{fig:pd}. The two vectors are well aligned for all regions in the universe that can be probed (i.e. are associated to Bisous filaments), as shown by the green line in each panel in Fig.~\ref{fig:pd}. This indicates that Bisous filaments appear to trace the cosmic velocity field well (and vice versa). Note that the two directions reported here originate from two vastly different data sets with different inherent biases: the Bisous filaments from galaxy redshift surveys and the V-web from reconstructions of the over-density and velocity field based on an opportunistic and incomplete compilation of galaxy distances using a variety of methods. Their close alignment indicates that these two completely independent methods are properly tracing the underlying distribution of matter.

Although by definition, all locations that can be tested for alignment are in Bisous filaments, these regions of the Universe are not necessarily defined as filaments by the shear analysis. Indeed, these locations have a range of densities and kinematic classifications. In total, 16.9\% of Bisous filament locations are V-web knots, 60.1\% are V-web filaments, 21.1\% are V-web sheets and 1.9\% are V-web voids (shown also in Fig.~\ref{fig:d}, top panel). These fractions can be contrasted with those found in the full volume to get a feeling for the nature of V-web defined environments that are sub-selected from the full reconstructed volume by the Bisous filaments. V-web filaments and knots are clearly favoured by the Bisous model -- these are over-represented (with respect to a random sampling of the V-web) by an extra 60\% and 80\% respectively. Sheets and voids are strongly unfavored, being seldomly selected by the Bisous model. In total, 77\% of Bisous filaments are either knots or filaments as defined by the V-web. These statistics indicate that the Bisous model and Shear tensor -- although not in perfect agreement -- agree well in terms of environment. It should also be added that this agreement occurs without any arbitrary adjustment to the threshold used to segregate different V-web environments.

The alignment between Bisous filaments and the \ethree~axes can thus be tested for environmental dependence by examining how it changes as a function of web type. The four panels in Fig.~\ref{fig:pd} show the alignment for void, sheet, filament and knot regions. The mean and median cosine of the distribution is also indicated in each panel and is many standard deviations away from the value expected from a uniform distribution ($\langle\cos(\theta)\rangle=0.5$). Although the alignment is statistically very significant in all four V-web environments, the two methods agree best in voids.  This is likely due to the fact that voids, being the most linear regions of the universe, are where the reconstructed velocity field most accurately represents the true velocity and over-density field.

The different environments (on scales of the Wiener Filter) probed by our web classification is examined in Fig.~\ref{fig:d} where we show the distribution of the sum of the eigenvalues ($\nabla\cdot v = \lambda_{\rm 1}+\lambda_{\rm 2}+\lambda_{\rm 3}$) for the Bisous filament locations tested in this study (top) and for the entire CF2 velocity field (bottom). The sum of the shear tensor's eigenvalues is the divergence of the velocity which on the linear scales of the Wiener Filter is proportional to the over-density. Both top and bottom panels of Fig.~\ref{fig:d} shows that knot regions (collapsing along all three principal directions) are the highest over-density while void regions (expanding along all three principal directions) are the lowest over-density environments. If the distribution of matter in the universe constituted a perfect Gaussian random field, these four curves would be Gaussian distributions,  whose standard deviation and mean increase as one goes from voids to knots. When the full CF2 volume within 100~Mpc is examined (Fig.~\ref{fig:d} bottom), the distribution of this quantity shows peaks in the sheet and filament web types. Note that these two curves are fairly wide indicating that within the volume considered here, web types overlap  in over-density (although the overlap is not extreme in the sense that for the most part overlap occurs only with adjacent web types.) A similar picture, but with different is seen in regions where the Bisous model is selected (Fig.~\ref{fig:d} top). However here the primacy of filaments and knots is clearly visible. In regions where the Bisous catalogue identifies filaments, these regions appear to be well sampled by the Wiener Filter reconstruction and are thus not {\it by construction} the mean:  as a result the eigenvalue sum distribution is considerably different than in the case of the full volume.

\section{Conclusions and Discussion}

There are many ways of looking at the large-scale distribution of matter. Traditionally this has been done by examining the spatial distribution of galaxies obtained primarily through large redshift surveys. Such an approach, although intuitive, is susceptible to well known issues related to redshift-space distortions such as fingers-of-god or the Kaiser effect. An alternative approach is to use the measured velocity field combined with reconstruction algorithms to infer full (luminous+dark) underlying 3D density field. Although attractive because it probes the full 3D density field, this approach too suffers from well known problems of accuracy and precision due to sparse sampling.

Nevertheless both methods can be used to study the so-called ``cosmic web'', namely the complex network of filaments and voids that connect galaxies in the Universe. Here, we have studied two such methods: the ``Bisous'' model based on a marked point process and the ``V-web'' a dynamical classification of the large-scale structure based on the shear field. The former, results in a filament catalogue based on the 2MRS galaxy catalogue while the latter uses a Wiener Filter to reconstruct the velocity field from the \mbox{Cosmicflows-2} survey of peculiar velocities. Both methods have been used to examine the properties of galaxies within the cosmic web \citep[see][]{2015MNRAS.448.1767C,2015MNRAS.446.1458M,2012MNRAS.421L.137L,2015arXiv150305915L,2015MNRAS.450.2727T,2015ApJ...800..112G}

The two methods, based on very different data sets and methods, agree well when it comes to identifying the principal directions of the large-scale structure. Fig.~\ref{fig:pd} shows that the spines of filaments from the Bisous model are well aligned with the axes that the V-web associates with filaments, \ethree. This success is not trivial, because of the very different limitations of each individual method. {\it A priori} one would hope, but not expect, that these two methods agree.

In fact, the agreement found here is considerably weaker than when the two methods are applied against each other in simulations, where biases and incompleteness play no role. Such a study was carried out by \cite{2014MNRAS.437L..11T} who found a near perfect alignment between \ethree~computed from the shear field and Bisous filaments (specifically, $\sim$80\% of filament spines were aligned within $30\degr$ of each other). When applied to data, the number of alignments within $30\degr$ is closer to 23\% -- still well above the 13.4\% one would expect for a random distribution, but considerably worse than in simulations. The reason the two methods fare poorer when applied to data is likely due to the Wiener Filtering reconstruction algorithm which smooths out the velocity field and perturbs its principal directions. Also, the disagreement may be a direct result of the fact that the two techniques are applied to drastically different data sets.

Perhaps the most important conclusion this study reaches, is a ``proof of concept''. Reconstruction methods based on the velocity field appear to do a good job in recovering the filamentary network of the cosmic web, albeit with large effective smoothing kernels. In that respect, as measurements of the cosmic velocity field progress in parallel with deeper galaxy redshift surveys, the agreement between the two methods is also likely to improve.

\section*{Acknowledgments}
NIL is supported by the Deutsche Forschungs Gemeinschaft. ET acknowledge the support from the ESF grants IUT26-2, IUT40-2. HC acknowledge support from the Lyon Institute of Origins under grant ANR-10-LABX-66.

\bibliography{Allrefs}

\label{lastpage}
\end{document}